\documentstyle[12pt]{article}
\begin{document}
\baselineskip6.8mm
\renewcommand{\theequation}{\thesection.\arabic{equation}}

\vskip 2cm
\title{Cosmic Strings and Black Holes${}^{\dag}$}
\author{\\
A.L. Larsen${}^{*}$}
\maketitle
\noindent
{\em
Theoretical Physics Institute, Department of Physics, \ University of
Alberta, Edmonton, Canada T6G 2J1}
\begin{abstract}
\baselineskip=1.5em
In the first part of this
talk, I consider some exact string solutions in curved spacetimes.
In curved spacetimes with a Killing vector (timelike or spacelike),
the string equations of motion
and constraints are reduced  to the Hamilton equations of a relativistic
point-particle in a scalar potential, by imposing a particular ansatz. As
special examples I consider circular strings in axially symmetric spacetimes,
as well as stationary strings in stationary spacetimes.

In the second part of the talk, I then consider in
more detail the stationary strings in the Kerr-Newman geometry. It is shown
that the world-sheet of a stationary string, that passes
the static limit of the 4-D
Kerr-Newman black hole, describes a 2-D black hole.
Mathematical results for
2-D black holes can therefore be applied to physical
objects;
(say) cosmic strings in the vicinity of Kerr black holes. As
an immediate
general result, it follows that the string modes are thermally excited.
\end{abstract}
\noindent
${}^{\dag}$Contribution to the proceedings of the Network-Meeting "String
Gravity", Paris, France, June  1996 and to the e-proceedings of the Workshop
"Non-Equilibrium Phase Transitions", Santa Fe, New Mexico, July 1996.
$^{*}$Electronic address: alarsen@phys.ualberta.ca
\newpage
\section{Introduction}
\setcounter{equation}{0}
In a generic curved spacetime, the classical equations of motion of
a bosonic string take the simplest form in the orthonormal gauge:
\begin{equation}
\ddot{x}^\mu-x''^\mu+\Gamma^\mu_{\rho\sigma}(\dot{x}^\rho\dot{x}^\sigma-
x'^\rho x'^\sigma)=0,
\end{equation}
supplemented by the constraints:
\begin{equation}
g_{\mu\nu}\dot{x}^\mu x'^\nu=g_{\mu\nu}(\dot{x}^\mu\dot{x}^\nu+x'^\mu
x'^\nu)=0.
\end{equation}
Here dot and prime denote differentiation with respect to the world-sheet
coordinates tau and sigma, respectively. $\Gamma^\mu_{\rho\sigma}$ is the
Christoffel symbol corresponding to the spacetime metric $g_{\mu\nu}.$

The main complication, as
compared to the case of flat Minkowski spacetime,
is related to the non-linearity
of the equations of motion (1.1). It makes it possible to obtain the complete
analytic solution only in a very few special cases like conical spacetime
\cite{san1} and plane-wave/shock-wave backgrounds \cite{san2}. There are
however
also very general results concerning integrability and solvability for
maximally symmetric spacetimes \cite{mic1,four} and gauged WZW models
\cite{bar,six}. These are the exceptional cases; generally the string equations
of motion in curved spacetimes are not integrable and even if they are, it
is usually an extremely difficult task to actually separate the equations,
integrate them and finally write down the complete solution in closed form.
Fortunately, there are several different ways to "attack" a system of coupled
non-linear partial differential equations, using either numerical,
approximative
or exact methods (for some recent reviews on strings in curved spacetimes,
see for instance \cite{veg1,all1}).

In this paper, we consider some exact string solutions in curved spacetimes,
obtained by imposing a particular ansatz. In Section 2, it is shown that in
curved spacetimes with a Killing vector (timelike or spacelike),
the string equations of motion
and constraints (1.1)-(1.2) reduce to the Hamilton equations of a relativistic
point-particle in a very simple scalar potential. As
special examples we consider circular strings in axially symmetric spacetimes
as well as stationary strings in stationary spacetimes. In Section 3, we
consider in
more detail the stationary strings in the Kerr-Newman geometry. It is shown
that the world-sheet of a stationary string, that passes
the static limit of the 4-D
Kerr-Newman black hole in the equatorial plane,
describes a 2-D Reissner-Nordstr\"{o}m black hole.
Mathematical results for
2-D black holes can therefore be applied to physical
objects;
(say) cosmic strings in the vicinity of Kerr black holes. An immediate
general result is that the string modes are thermally excited.
\section{Exact String Solutions}
\setcounter{equation}{0}
In this section we shall consider some exact string solutions in curved
spacetimes. We first consider, as special examples, circular strings and
stationary strings. Eventually we then give a more general description valid in
curved spacetimes with an arbitrary Killing vector.
\vskip 10pt
\hspace*{-6mm}{\bf A. Circular Strings}\\
The condition for a curved spacetime to allow for circular strings, is that the
spacetime is axially symmetric. We can thus use a coordinate system
$x^\mu=(t,r,\theta,\phi)$ where the metric is explicitly independent of the
azimuthal angle $\phi:$
\begin{equation}
ds^2=g_{\mu\nu}(t,r,\theta)dx^\mu dx^\nu.
\end{equation}
We consider for simplicity a 4-D spacetime, since we have mainly cosmic
strings in mind. The ansatz which describes a dynamical circular string is
provided by:
\begin{equation}
t=t(\tau),\;\;\;\;\;\;r=r(\tau),\;\;\;\;\;\;\theta=\theta(\tau),\;\;\;\;\;\;
\phi=\sigma+f(\tau).
\end{equation}That is, we identify the spatial world-sheet coordinate $\sigma$
with the azimuthal angle $\phi$ up to the addition of a function of the
world-sheet coordinate $\tau,$ and let the other coordinates depend only on
$\tau;$ this guarantees that the string is circular. The four functions
$t(\tau),\;r(\tau),\;\theta(\tau)$ and $f(\tau)$ are now to be
determined by the equations of motion and constraints (1.1)-(1.2).

The invariant definition of a circular string in an axially symmetric spacetime
is that the Killing vector corresponding to the axial symmetry is tangent to
the
string world-sheet. It is then straightforward to show that the ansatz (2.2) is
in fact the most general ansatz describing circular strings, at least up to
residual gauge transformations and constant rescalings of $\sigma.$

By inserting the ansatz (2.2) into eqs.(1.1), it can be shown that the
equations of motion reduce to the Hamilton equations of the point-particle
Hamiltonian:
\begin{equation}
{\cal H}=\frac{1}{2}g^{\mu\nu}{\cal P}_\mu{\cal P}_\nu+\frac{1}{2}g_{\phi\phi},
\end{equation}
while the two constraints eqs.(1.2) become:
\begin{equation}
{\cal H}=0,\;\;\;\;\;\;\;\;{\cal P}_{\phi}=0.
\end{equation}
The dynamics of a circular string in an axially symmetric curved spacetime
is thus mathematically equivalent to
the dynamics of a {\it massless} zero-angular-momentum
point-particle, moving in the same curved spacetime as well as
in a scalar potential given simply by the $(\phi\phi)$-component
of the metric. By a conformal transformation, this system is also
equivalent to the
system of a {\it massive} point-particle moving in some unphysical spacetime
(see \cite{all2} for a discussion of this point in the more general case of
charge-current carrying circular strings, as well as the original references
\cite{fro,car}
for the similar discussion for stationary strings. See also Section 3);
however, the latter
approach will not be considered in this section. We find it more convenient to
work
directly with the physical metric $g_{\mu\nu}.$

It must be stresset that the reduction of the system (1.1)-(1.2) to the system
(2.3)-(2.4) of course does not mean that we have solved anything. In fact,
Hamiltonian systems of the form (2.3), with or without the scalar potential
$g_{\phi\phi},$ are generally not separable, and even if the "standard"
point-particle part $g^{\mu\nu}{\cal P}_\mu{\cal P}_\nu$ of the Hamiltonian is
separable, the separability is generally destroyed by the scalar potential
$g_{\phi\phi}.$ This is precisely what happens in the case of the Kerr-Newman
black hole: the $g^{\mu\nu}{\cal P}_\mu{\cal P}_\nu$-part of the Hamiltonian
(2.3) is separable, as is well known, but the $g_{\phi\phi}$-part is not of
separable type \cite{all2,all3}.

In most cases it is therefore necessary to make further reductions. Consider
for instance 4-D spacetimes of the form:
\begin{equation}
ds^2=g_{tt}(r,\theta)dt^2+g_{rr}(r,\theta)dr^2+g_{\theta\theta}
(r,\theta)d\theta^2+g_{\phi\phi}(r,\theta)d\phi^2+
2g_{t\phi}(r,\theta)dt d\phi,
\end{equation}
which covers the black hole spacetimes, some cosmological spacetimes
(in static coordinates) the $2+1$ black hole and black string spacetimes
(taking
$\theta=const.$) etc. If we further consider circular strings in the equatorial
plane $\theta=\pi/2,$
then it can be shown that the equation determining the radius of the
string-loop, as obtained from the Hamiltonian (2.3), takes the form:
\begin{equation}
\dot{r}^2+V(r)=0,
\end{equation}
where:
\begin{equation}
V(r)=g^{rr}(E^2g^{tt}+g_{\phi\phi});\;\;\;\;\;\;\;\;E=const.
\end{equation}
That is, everything is solved in quadratures, and the dynamics of the circular
string is exactly known in closed form.

Using this approach, it is now a relatively
easy task to describe the circular string
dynamics in any curved spacetime of the form (2.5). Simply by looking at the
line-element, one can immediately read off the potential $V(r),$ which
determines the dynamics of the circular string. Next, one can then solve
the equations exactly and the physical properties (energy-density, pressure,..)
can be extracted.

Circular strings were first discussed in Minkowski spacetime in \cite{vil}.
More recently the dynamics of circular strings
has been investigated in detail in both cosmological and black hole
spacetimes [14-26] (charged circular strings in curved spacetimes were
discussed in \cite{all3} and references given therein).
We refer the interested reader to these publications
and only give a couple of examples here. In general the
dynamics of a circular string is
governed by the tension, which always tries to contract the string, and the
local gravitational field which can be either attractive or repulsive.

In the
absence of a repulsive gravitational field, the circular string will always
collapse from some maximal size. This is illustrated in Fig.1 in the cases of
(a) Minkowski, (b) Schwarzschild, (c) anti de Sitter and (d) Schwarzschild-anti
de Sitter. In all these cases the string collapses to zero size. There are
however differences due to the causal structure of the spacetimes. In
Minkowski and anti de Sitter spacetimes, the string dynamics is truely
oscillatory (we neglect the gravitational backreaction), while in the black
hole spacetimes the dynamics stops when the string falls into the
spacetime singularity.

In the presence of a repulsive gravitational field, the situation is more
complicated. A circular string can be oscillating between a maximal size
and zero size, oscillating between a maximal size and a non-zero minimal size,
it can be expanding forever or it can even be of constant size. This is
illustrated in Fig.2 in the cases of (a) de Sitter and (b) Kerr black hole.
The potential shown for de Sitter spacetime corresponds to the special
case where
$H^2E^2<1/4,$ where $H$ is the Hubble constant and $E$ is the integration
constant introduced in (2.7); the more general analysis for arbitrary
values of $H^2E^2$ is presented in
\cite{veg2}.
{}From the potential (a) follows that small strings will be oscillating
between a maximal size and zero size, while large strings will bounce on
the barrier and re-expand
forever. The other types of circular string dynamics in
de Sitter spacetime are described in detail in \cite{veg2}.

The potential shown for the Kerr Black hole corresponds to the special case
where $a^2>E^2,$ where $a$ is the angular momentum of the black hole;
the more general potential for arbitrary
values of $a$ is given in
\cite{all5}.
{}From the potential (b) follows that a circular string contracts from
its maximal size to a non-zero minimal size. When the string reaches its
minimal size, it is actually inside the event horizon and thus cannot re-expand
because of the causal structure. However, such solution can be interpreted
as a string contracting in one spacetime, expanding in another etc
\cite{ini,all5}.
\vskip 10pt
\hspace*{-6mm}{\bf B. Stationary Strings}\\
As another example of a family of exact string solutions in curved spacetimes,
we now consider stationary strings. It is well known that in flat Minkowski
spacetime there is only one stationary string configuration, namely the
straight string; a string with any other shape will start vibrating. However,
we shall now show that the situation in curved spacetimes is completely
different. In fact, in some curved spacetimes stationary strings can have
infinitely many completely
different shapes, and there can be very interesting physics
associated with them.

The condition for a curved spacetime to allow for stationary
strings, is that the
spacetime itself is stationary. We can thus use a coordinate system
$x^\mu=(t,r,\theta,\phi)$ where the metric is explicitly independent of the
time-coordinate $t:$
\begin{equation}
ds^2=g_{\mu\nu}(r,\theta,\phi)dx^\mu dx^\nu.
\end{equation}
The ansatz which describes a stationary string is:
\begin{equation}
t=\tau+f(\sigma),\;\;\;\;\;\;
r=r(\sigma),\;\;\;\;\;\;\theta=\theta(\sigma),\;\;\;\;\;\;
\phi=\phi(\sigma).
\end{equation}That is, we identify the temporal world-sheet coordinate $\tau$
with the time-coordinate $t$ up to the addition of a function of the
world-sheet coordinate $\sigma,$ and let the other coordinates depend only on
$\sigma;$ this guarantees that the string is stationary. The four functions
$f(\sigma),\;r(\sigma),\;\theta(\sigma)$ and $\phi(\sigma)$ are now to be
determined by the equations of motion and constraints (1.1)-(1.2).

C.f. the previous discussion of circular strings in axially symmetric
spacetimes,
the invariant definition of a stationary string in a
stationary spacetime
is that the timelike Killing vector $\partial_{t}$ is tangent to the
string world-sheet. It is then straightforward to show that the ansatz (2.9) is
the most general ansatz describing stationary strings up to
residual gauge transformations and constant rescalings of $\tau.$

By inserting the ansatz (2.9) into eqs.(1.1), it can be shown that the
equations of motion reduce to the Hamilton equations of the point-particle
Hamiltonian:
\begin{equation}
{\cal H}=\frac{1}{2}g^{\mu\nu}{\cal P}_\mu{\cal P}_\nu+\frac{1}{2}g_{tt},
\end{equation}
while the two constraints eqs.(1.2) become:
\begin{equation}
{\cal H}=0,\;\;\;\;\;\;\;\;{\cal P}_{t}=0,
\end{equation}
compare with eqs.(2.3)-(2.4). Thus mathematically the problem of finding the
stationary string configurations looks very similar to the problem of
describing the dynamical circular strings.
It is however a very interesting result that in the case of the Kerr-Newman
black hole, the Hamiltonian system (2.10) is actually separable \cite{fro}.
Thus
contrary to the case of the circular strings in the Kerr-Newman background,
the $g_{tt}$-part of the
Hamiltonian (2.10)
does not destroy the separability of the system, and therefore
the stationary string configurations can be described
completely and explicitly \cite{fro}.

In most other cases it is however still
necessary to make further reductions to solve the system. If we consider
again the 4-D spacetimes of the form (2.5), and restrict ourselves to
describing
only stationary strings in the equatorial
plane $\theta=\pi/2,$
then it can be shown that the equation determining the radial coordinate of the
stationary string, as obtained from the Hamiltonian (2.10), takes the form:
\begin{equation}
r'^2+U(r)=0,
\end{equation}
where:
\begin{equation}
U(r)=g^{rr}(L^2g^{\phi\phi}+g_{tt});\;\;\;\;\;\;\;\;L=const.
\end{equation}
That is, everything is solved in quadratures, and the
stationary string configuration is exactly known in closed form.

Using this approach, it is now a relatively
easy task to describe the stationary strings
in any curved spacetime of the form (2.5). Everything goes exactly as in the
previous case of the circular strings: simply by looking at the
line-element, one can immediately read off the potential $U(r),$ which
determines the location of the stationary string. Next, one can then solve
the equations exactly and the physical properties (energy-density, pressure,..)
can be extracted.

As already mentioned, the only stationary string in Minkowski spacetime
is the straight string. In stationary spacetimes with an attractive
gravitational field, the stationary strings typically extend out to spatial
infinity. This is illustrated in Fig.3 in the cases of (a) Minkowski, (b)
Schwarzschild and (c) anti de Sitter spacetimes.  For the potentials
shown, the two ends of the stationary string are at spatial infinity, and
the string passes $r=0$ with some impact parameter related to the
integration constant $L.$ For a more complete discussion of these stationary
strings we refer the reader to \cite{fro,all8}.

If the spacetime has a repulsive
gravitational field it is also possible to have stationary strings in a
small compact region. This is illustrated in Fig.4 in the case of de Sitter
spacetime. In this case the stationary string is located somewhere between
two finite values of the radial coordinate (both being smaller than the
horizon size). It turns out that there are infinitely many completely
different stationary string configurations in de Sitter spacetime classified,
using a simple underlying mathematical structure, in terms of two integers
\cite{all8}.
An example of a stationary string in de Sitter spacetime is presented
in Fig.5. The complete discussion is given in \cite{all8}.
\vskip 10pt
\hspace*{-6mm}{\bf C. General Formalism}\\
It is appearent from the previous discussions that the mathematics of the
circular strings is very similar to the mathematics of the stationary
strings, although the physics is of course completely different. By
comparing the formulas (2.1)-(2.7) with (2.8)-(2.13) we notice that there
is actually a kind of "duality" between circular strings in axially symmetric
spacetimes and stationary strings in stationary spacetimes \cite{all2,all8};
the
"duality-transformation" being simply:
\begin{equation}
t\;\longleftrightarrow\;\phi\;,\;\;\;\;\;\;\;\;
\tau\;\longleftrightarrow\;\sigma
\end{equation}
It must be stressed, however, that there is absolutely nothing mysterious
about this "duality". In fact, the previous discussions of circular strings
and stationary strings are just special examples of a more general approach
for obtaining special exact solutions of the string equations of motion and
constraints in curved spacetimes with a Killing vector:\\
Consider an arbitrary curved spacetime with an arbitrary Killing vector
$\xi.$ Then take a coordinate system such that the Killing vector has the
standard form:
\begin{equation}
\xi^\mu=\delta^\mu_{i}\;,\;\;\;\;\;\mbox{for some}\;\;i.
\end{equation}
Now make the following ansatz for the string:
\begin{equation}
x^{i}=\sigma^1+f(\sigma^2),\;\;\;\;\;\;\;\;x^{j}=x^{j}(\sigma^2);\;\;\;j
\neq i
\end{equation}
where $(\sigma^1,\sigma^2)$ are the two world-sheet coordinates. That is,
the targetspace coordinate corresponding
to the Killing vector is identified with
one of the world-sheet coordinates plus a function of the other, while the
other
targetspace coordinates are only functions of the other world-sheet coordinate.
In that case it is easy to show that the string equations of motion and
constraints (1.1)-(1.2) reduce to the Hamiltonian system:
\begin{equation}
{\cal H}=\frac{1}{2}g^{\mu\nu}{\cal P}_\mu{\cal P}_\nu+\frac{1}{2}g_{ii},
\end{equation}
\begin{equation}
{\cal H}=0,\;\;\;\;\;\;\;\;{\cal P}_{i}=0.
\end{equation}
This is just the Hamiltonian of a massless point-particle moving in the
curved spacetime as well as in a scalar potential given simply by the
$(ii)$-component of the metric.
If the Killing vector is timelike, this system describes stationary
strings, while if it is spacelike, it describes
dynamical strings of some non-trivial shape.
\section{Stationary Strings and 2-D Black Holes}
\setcounter{equation}{0}
(This section is essentially a reproduction of [28])\\
The aim of this section is to attract the attention to
the
remarkable fact, that the study of propagation of perturbations along
a
stationary cosmic string located in the gravitational field of a stationary
4-dimensional black hole, is directly related with physics of
2-dimensional
black holes.
If one neglects the gravitational effects of a cosmic string and
assumes that its thickness is zero, then a string configuration in a
given
gravitational field is  a minimal surface, and its equations of
motion
can be obtained
by variation of the Goto-Nambu action; these equations are of course equivalent
to eqs.(1.1)-(1.2). An important case is when both
the
gravitational field and a string configuration are stationary. This
case
corresponds to a physical situation of a string which is in
equilibrium
in the gravitational field.
The problem of finding the equilibrium string
configurations in a
stationary spacetime can be reduced to the problem of solving
geodesic
equations in a 3-dimensional unphysical
space \cite{fro}. The remarkable property
of the gravitational field of a stationary black hole is that these
geodesic
equations can be solved in quadratures \cite{fro,car}.
Consider a 2-dimensional  time-like world-sheet $\Sigma$ of
the string defined by the equations  $x^{\mu}=x^{\mu}(\sigma^{A})$
($x^{\mu}$ ($\mu =0,...,4$) denote  spacetime coordinates and
$\sigma^{A}$ ($A=0,1$) are the coordinates on the world-sheet).
The metric  $G_{AB}$  induced on $\Sigma$ reads:
\begin{equation}
G_{AB}=g_{\mu \nu}x^{\mu}_{,A}x^{\nu}_{,B} .\label{1}
\end{equation}
where $g_{\mu\nu}$ is the spacetime metric. Consider a stationary
spacetime and
let $\xi^{\mu}$ be the corresponding
4-dimensional Killing vector. We call a string stationary if
$\xi^{\mu}$ is
tangent to its world sheet $\Sigma$. Denote by $\eta^{A}$ a
2-dimensional
vector $\eta^{A}=G^{AB}x^{\mu}_{,B}\xi_{\mu}$. It is easy to show
that:
\begin{equation}
\xi^{\mu}=\eta^{A}x^{\mu}_{,A},
\end{equation}
\begin{equation}
\eta_{A|B}=x^{\mu}_{,A}x^{\nu}_{,B}\xi_{\mu;\nu} ,
\end{equation}
where semicolon and vertical line denote covariant derivatives
with respect to 4 and 2-dimensional metrics, respectively. The last
relation implies that $\eta$ is a Killing
vector for the induced metric $G_{AB}$.
Denote:
\begin{equation}
F=-g_{\mu\nu}\xi^{\mu}\xi^{\nu}
=-G_{AB}\eta^{A}\eta^{B}.
\end{equation}
Then the
induced line-element $dl^2=G_{AB}d\sigma^{A}
d\sigma^{B}$ on $\Sigma$ can be written as:
\begin{equation}
dl^2=-Fd\hat{\tau}^2+F^{-1}d\hat{\sigma}^2,\label{2}
\end{equation}
where $\hat{\tau}=\hat{\tau}(\sigma^{A}),\;\;
\hat{\sigma}=\hat{\sigma}(\sigma^{A}).$
This representation is valid in the regions where $F\ne 0$, and hence
$\xi$ is
either time-like or space-like.
We assume that the 4-dimensional spacetime is asymptotically flat and
contains
a black hole. We assume also that
$\hat{\sigma}=\infty$ corresponds to the points of the string located
in the
asymptotically flat region of the physical spacetime,
so that $F(\hat{\sigma}=\infty)=1$. The metric, eq.(3.5),
describes a 2-dimensional black hole if $F=0$ at finite value of
$\hat{\sigma}$. At this
point the Killing 2-vector $\eta$ is null. It happens at the points
where the
string world-sheet crosses the infinite-red-shift surface
(the static limit), i.e. the surface where $\xi^2=0$. For a static
black hole
this surface coincides with the event horizon. For stationary
(rotating) black holes it is lying outside the horizon.
The region located between infinite-red-shift surface and
the event
horizon of a stationary black hole is known as the ergosphere.
Points of the string
located inside the ergosphere thus
correspond to the interior of the 2-dimensional
black hole.

The main observation we would like to make now is that perturbations
propagating along the string can be described by a coupled system of
a pair of scalar field equations in the 2-dimensional metric
$G_{AB}$ \cite{all}.
For this reason, if a stationary string is passing through the
ergosphere
of a 4-dimensional black hole, the physics of string excitations is
effectively
reduced to  the physics of 2-dimensional black holes.

We shall first derive the differential equations describing the
perturbations
propagating along a stationary string configuration embedded in an
arbitrary
stationary 4-dimensional spacetime. We then consider, as a special
example,
stationary strings in the background of a Kerr-Newman black hole.
\vskip 12pt
\hspace*{-6mm}To be more specific, we write the metric of a generic
4-dimensional
stationary spacetime in the form:
\begin{equation}
g_{\mu\nu}=\left( \begin{array}{cc} -F & -FA_{i}\\
-FA_{i} & -FA_{i}A_{j}+H_{ij}/F\end{array}\right),
\end{equation}
where $\partial_t F=0,\;\partial_t A_{i}=0,\;\partial_t H_{ij}=0.$
That is to say,
the Killing vector is given explicitly by:
\begin{equation}
\xi^\mu=(1,\;0,\;0,\;0),\;\;\;\;\;\;\xi_\mu=(-F,\;-FA_{i}),
\end{equation}
consistent with the notation of eq.(3.4).
A stationary string configuration is parametrized in the following
way:
\begin{equation}
t=x^0=\tau,\;\;\;\;\;x^{i}=x^{i}(\sigma){\;};\;\;\;\;\;\;\;\;
(\tau\equiv\sigma^0,\;\sigma\equiv\sigma^1).
\end{equation}
Then the equations of motion corresponding to the Goto-Nambu action
reduce
to
\cite{fro}:
\begin{equation}
x^{i}{''}+\tilde{\Gamma}^{i}_{jk}x^{j}{'}
x^{k}{'}=0,\;\;\;\;\;\;H_{ij}x^{i}{'}x^{j}{'}=1,
\end{equation}
where $\tilde{\Gamma}^{i}_{jk}$ is the Christoffel connection for the
metric $H_{ij}$ and a prime denotes differentiation with respect to
$\sigma.$ The induced metric on the world-sheet now takes the form:
\begin{equation}
G_{AB}=\left( \begin{array}{cc} -F & -FA\\
-FA & -FA^2+1/F\end{array}\right);\;\;\;\;\;\;\;\;A\equiv
A_{i}x^{i}{'},
\end{equation}
so that $\mbox{det}\ G=-1$.
The following coordinate transformation on the world-sheet:
\begin{equation}
\hat{\tau}=\tau+\int^{\sigma}A\;d\sigma,\;\;\;\;\;\;\;\;\hat{\sigma}=\sigma,
\end{equation}
brings the induced line element into the form of eq.(3.5), that is:
\begin{equation}
\hat{G}_{AB}=\left( \begin{array}{cc} -F & 0\\
0 & 1/F\end{array}\right).
\end{equation}

A covariant approach describing the propagation of perturbations
along
an arbitrary string configuration embedded in an arbitrary
spacetime, was developed  in Ref.\cite{all}
(see also
\cite{car2,guv}). The
general transverse (physical) perturbation around a background
Goto-Nambu string configuration is written as:
\begin{equation}
\delta x^\mu=\Phi^R n^\mu_R,\;\;\;\;\;\;(R=2,3),
\end{equation}
where the two vectors $n^\mu_R,\;$ normal to the string world-sheet,
fulfill:
\begin{equation}
g_{\mu\nu}n^\mu_R n^\nu_S=\delta_{RS},\;\;\;\;\;\;\;\;\;\;g_{\mu\nu}
x^\mu_{,A}n^\nu_R=0,
\end{equation}
as well as the completeness relation:
\begin{equation}
g^{\mu\nu}=G^{AB}x^\mu_{,A} x^\nu_{,B}+\delta^{RS}n^\mu_R n^\nu_S.
\end{equation}
It can be shown \cite{all} that the perturbations $\Phi^R$
are determined by the following effective action:
\begin{equation}
{\cal S}_{\mbox{eff.}}=\int_{}^{}
d^2\zeta\sqrt{-G}\;\Phi^R\left\{G^{AB}
(\delta^T_R\nabla_A+\mu_R\hspace*{1mm}^T\hspace*{1mm}_A)
(\delta_{TS}\nabla_B+\mu_{TSB})+
{\cal V}_{RS}\right\}\Phi^S,
\end{equation}
where ${\cal V}_{RS}={\cal V}_{(RS)}$ are scalar potentials
and $\mu_{RS}\hspace*{1mm}^A=\mu_{[RS]}\hspace*{1mm}^A$
are vector potentials on
$\Sigma,$
determined by its embedding into the 4-dimensional spacetime,
and  $\nabla_A$ is the covariant derivative with
respect to  the metric $G_{AB}$.  The vector potentials
$\mu_{RSA}$
coincide with  the normal fundamental form:
\begin{equation}
\mu_{RSA}=g_{\mu\nu}n^\mu_R x^\rho_{,A}\nabla_\rho n^\nu_S,
\end{equation}
where $\nabla_\rho$ is the covariant derivative with respect to
the metric $g_{\mu\nu}$. The scalar potentials ${\cal V}_{RS}$
are defined as:
\begin{equation}
{\cal V}_{RS}\equiv\Omega_{RAB}\Omega_S\hspace*{1mm}^{AB}-
G^{AB}x^\mu_{,A}x^\nu_{,B}R_{\mu\rho\sigma\nu}n^\rho_R
n^\sigma_S,
\end{equation}
where $\Omega_{RAB}$ is the second fundamental form:
\begin{equation}
\Omega_{RAB}=g_{\mu\nu}n^\mu_R x^\rho_{,A}\nabla_\rho x^\nu_{,B},
\end{equation}
and $R_{\mu\rho\sigma\nu}$ is the Riemann tensor corresponding to
the metric $g_{\mu\nu}$.
The equation describing the propagation of  perturbations along the
string world-sheet is:
\begin{equation}
\left\{ G^{AB}(
\delta^T_R {\nabla}_A+{\mu}_R\hspace*{1mm}^T
\hspace*{1mm}_A)(
\delta_{TS}{\nabla}_B+{\mu}_{TSB})
+{{\cal V}}_{RS}\right\}\Phi^S=0.
\end{equation}

We obtain now explicit expressions for vector
${\mu}_{RS}\hspace*{1mm}^A$ and
scalar ${\cal V}_{RS}$ potentials which enter the propagation
equations for
a stationary string obeying eqs.(3.9), in the background (3.6).
Using eqs.(3.14), the normal vectors $n^\mu_R$ take the form:
\begin{equation}
n^\mu_R=(-A_{i}n^{i}_R,\;n^{i}_R),\;\;\;\;\;\;n_{\mu
R}=(0,\;F^{-1}H_{ij}
n^{j}_R)\equiv (0,\;n_{iR}).
\end{equation}
It is convenient to introduce also  3-dimensional vectors
\begin{equation}
\tilde{n}^{i}_R=|F|^{-1/2}n^{i}_R,
\end{equation}
which together with $x^{i}{'}$
form an  orthonormal
system $(x^{i}{'},\tilde{n}^{i}_2,\tilde{n}^{i}_3)$ in the
3-dimensional
space with metric $H_{ij}$.
We note that there is an ambiguity
$\tilde{n}^{i}_R\rightarrow \tilde{n}^{i}_R+\delta(\tilde{n}^{i}_R)$
in the choice of pair of normal
vectors  $\tilde{n}^{i}_R:$
\begin{equation}
\delta(\tilde{n}^{i}_R)=
\Lambda_R\hspace*{1mm}^S\tilde{n}^{i}_S\;
;\;\;\;\;\;\;\Lambda_{RS}(\zeta)=-\Lambda_{SR}(\zeta).
\end{equation}
Under these 'gauge'  transformations, which leave
invariant the perturbations (3.13) and the effective action (3.16),
the quantities $\Phi^S$, ${\mu}_{RSA}$, and ${\cal V}_{RS}$
are transformed as follows:
\begin{eqnarray}
\delta(\Phi^R)\hspace*{-2mm}&=&\hspace*{-2mm}
\Lambda^R\hspace*{1mm}_S\Phi^S,\nonumber\\
\delta(\mu_{RSA})
\hspace*{-2mm}&=&\hspace*{-2mm}\Lambda_R\hspace*{1mm}^T\mu_{TSA}
-\Lambda_S\hspace*{1mm}^T\mu_{TRA}-\Lambda_{RS,A},\\
\delta({{\cal V}}_{RS})\hspace*{-2mm}&=&\hspace*{-2mm}
\Lambda_R\hspace*{1mm}^T{{\cal V}}_{TS}+
\Lambda_S\hspace*{1mm}^T{{\cal V}}_{RT} .\nonumber
\end{eqnarray}
We fix
the freedom of  'gauge'  by
choosing the vectors $\tilde{n}^{i}_R$ to be covariantly constant in
the 3-dimensional  space \cite{all}.
After straightforward but tedious
calculations we get:
\begin{equation}\label{mu}
\mu_{RS}\hspace*{2mm}^{A}=\eta^A\;\frac{A_{ij}}{2}
n^{i}_R n^{j}_S=\eta^{A}\;\frac{1}{2F^2}\xi_{[0,i}\xi_{j]}
n^{i}_R n^{j}_S,
\end{equation}
\begin{eqnarray}
{\cal
V}_{RS}\hspace*{-2mm}&=&\hspace*{-2mm}-x^{i}{'}x^{j}{'}\tilde{R}_{iklj}
n^{k}_R
n^{l}_S+\frac{\delta_{RS}}{2}x^{i}{'}x^{j}{'}(\tilde{\nabla}_{i}
\tilde{\nabla}_{j}F)\nonumber\\
\hspace*{-2mm}&-&\hspace*{-2mm}\frac{\delta_{RS}}{4F}x^{i}
{'}x^{j}{'}(\tilde{\nabla}_{i}F)(\tilde{\nabla}_{j}F)
+\frac{F}{4}\delta^{TU}A_{il}A_{kj}n^{i}_U n^{j}_T n^{k}_R n^{l}_S.
\end{eqnarray}
Here the "field-strength" $A_{ij}$ is defined by:
\begin{equation}
A_{ij}=A_{i,j}-A_{j,i}=\tilde{\nabla}_{j}A_{i}-\tilde{\nabla}_{i}A_{j},
\end{equation}
where $\tilde{\nabla}_{i}$ is the covariant derivative with respect
to the
metric $H_{ij}$
and $\tilde{R}_{iklj}$ is the Riemann tensor corresponding to the
metric $H_{ij}.$ It is easy to verify that for our choice of 'gauge'
the
vector potentials $\mu_{RS}\hspace*{2mm}^{A}$ obey the analog
of the Lorentz
gauge conditions  $\nabla_A\mu_{RS}\hspace*{1mm}^{A}=0$.
The anti-symmetric products of normal vectors, appearing in the
scalar and vector potentials, eqs.(3.25)-(3.26),
can be eliminated using the identity:
\begin{equation}
\epsilon^{RS}\tilde{n}^{i}_R\tilde{n}^{j}_S=\tilde{e}^{ijk}H_{kl}x^{l}
{'};
\;\;\;\;\;\;\;\;\tilde{e}^{ijk}=(H)^{-1/2}\epsilon^{ijk}.
\end{equation}
In particular we have:
\begin{equation}
x^{i}{'}x^{j}{'}\tilde{R}_{iklj}n^{k}_R n^{l}_S=
F\delta_{RS}(\frac{1}{2}\tilde{R}-\tilde{R}_{ij}x^{i}{'}
x^{j}{'})-\tilde{R}_{ij}n^{i}_R n^{j}_S,
\end{equation}
\begin{equation}
\frac{F}{4}\delta^{TU}A_{il}A_{kj}n^{i}_U n^{j}_T n^{k}_R n^{l}_S=
\frac{F^3}{4}\delta_{RS}(A_{ik}A_{j}\;^{k}x^{i}{'}x^{j}{'}-\frac{1}{2}
A_{ij}
A^{ij}).
\end{equation}
By using the equations (3.29)-(3.30) we get:
\begin{equation}
\mu_{RS}\hspace{1mm}^A=\mu \epsilon_{RS}\eta^A\;\; ;\;\;
\;\;\;\;\;\ \mu=\frac{F}{4}A_{ij}
\tilde{e}^{ijk}H_{kl}x^{l}{'},
\end{equation}
\begin{eqnarray}
{\cal  V}_{RS}\hspace{-2mm}&=&\hspace{-2mm}\delta_{RS}
[\frac{x^{i}{'}x^{j}{'}}{2}(\tilde{\nabla}_{i}\tilde{\nabla}_{j}F)-
\frac{1}{4F}x^{i}
{'}x^{j}{'}(\tilde{\nabla}_{i}F)(\tilde{\nabla}_{j}F)-
F(\frac{1}{2}\tilde{R}-\tilde{R}_{ij}x^{i}{'}
x^{j}{'})\nonumber\\
\hspace{-2mm}&+&\hspace{-2mm}
\frac{F^3}{4}(A_{ik}A_{j}\;^{k}x^{i}{'}x^{j}{'}-\frac{1}{2}
A_{ij}A^{ij})]+\tilde{R}_{ij}n^{i}_R n^{j}_S .
\end{eqnarray}

As a special application of the above formalism, we now
consider
perturbations propagating
along a stationary string in the Kerr-Newman black hole
background. The Kerr-Newman metric in  Boyer-Lindquist coordinates
\cite{boy} reads:
\begin{equation}\label{BL}
ds^2=-\frac{\Delta}{\rho^2} [dt -a\sin^2\theta d\phi ]^2
+\frac{\sin^2\theta}{\rho^2}
[(r^2+a^2)d\phi -a dt]^2 +\frac{\rho^2}{\Delta}dr^2 +\rho^2 d\theta^2
,
\end{equation}
where  $\Delta=r^2-2Mr+Q^2+a^2$ and $\rho^2=r^2+a^2\cos^2\theta$.
This metric  is of the form  (3.6) with:
\begin{eqnarray}
H_{rr}=\frac{\Delta-a^2\sin^2\theta}{\Delta},
\hspace{.6cm}
H_{\theta\theta}=\Delta-a^2\sin^2\theta,
\hspace{.6cm}
H_{\phi\phi}=\Delta\sin^2\theta,
\end{eqnarray}
\begin{equation}
F=\frac{\Delta-a^2\sin^2\theta}{r^2+a^2\cos^2\theta},
\hspace{.6cm}
A_\phi=a\sin^2\theta\;\frac{2Mr-Q^2}{\Delta-a^2\sin^2\theta}.
\end{equation}
The unperturbed stationary string configurations are obtained by
solving
eqs.(3.9), \cite{fro}:
\begin{eqnarray}\label{30}
(H_{rr}r')^2\hspace*{-2mm}&=&\hspace*{-2mm}\frac{a^2 b^2}{\Delta^2}-
\frac{q^2}{\Delta}+1,\nonumber\\
(H_{\theta\theta}\theta')^2\hspace*{-2mm}&=&\hspace*{-2mm}q^2-
\frac{b^2}{\sin^2\theta}-a^2\sin^2\theta \equiv q^2-U(\theta),\\
(H_{\phi\phi}\phi')^2\hspace*{-2mm}&=&\hspace*{-2mm}b^2,\nonumber
\end{eqnarray}
where $b$ and $q$ are integration constants. Equations (\ref{30}) are
evidently invariant with respect to the reflection $\phi\rightarrow
-\phi$.
It means that if $(r(\sigma), \theta(\sigma), \phi (\sigma) )$ is
a
solution of eqs.(\ref{30}) then $(r(\sigma), \theta(\sigma), -\phi
(\sigma) )$
is also a solution.

We  consider at first the case when a stationary
string is located in the equatorial plane $\theta=\pi/2,$
corresponding to $|b|\geq |a|$ and $q^2=a^2+b^2.$
In that case, the two
covariantly constant normal vectors $\tilde{n}^{i}_R,$ introduced in
eq.(3.22), are given by:
\begin{equation}
\tilde{n}^{i}_2\equiv\tilde{n}^{i}_\perp=\frac{1}{\sqrt{|r^2-2Mr+Q^2|}}
(0,\;1,\;0),
\end{equation}
\begin{equation}
\tilde{n}^{i}_3\equiv\tilde{n}^{i}_\parallel=\frac{1}{\sqrt{|r^2-2Mr+Q
^2|}}
(-b,\;0,\;H_{rr}r').
\end{equation}
It is now straightforward to compute the vector and scalar potentials
given by eqs.(3.31),(3.32):
\begin{equation}
\mu=0,
\end{equation}
\begin{eqnarray}
{\cal V}_{\perp\perp}\hspace*{-2mm}&=&\hspace*{-2mm}
{\cal
V}_{\parallel\hspace*{1mm}\parallel}+\frac{2(M^2-Q^2)(a^2-b^2)}
{r^2(\Delta-a^2)^2}\nonumber\\
\hspace*{-2mm}&=&\hspace*{-2mm}\frac{M^2-Q^2}{r^2(\Delta-a^2)}
[1+\frac{2(a^2-b^2)}{\Delta-a^2}]+\frac{(r-M)(b^2-a^2)}{(\Delta-a^2)^2}
[\frac{M}{r^2}-\frac{Q^2}{r^3}]\nonumber\\
\hspace*{-2mm}&+&\hspace*{-2mm}\frac{\Delta-b^2}{(\Delta-a^2)^2}
[-\frac{2M}{r}+\frac{3Q^2}{r^2}+\frac{3M^2}{r^2}-\frac{6MQ^2}{r^3}
+\frac{2Q^4}{r^4}].
\end{eqnarray}

In the generic case, $|b|>|a|,$ the stationary configuration
in the equatorial plane describes
an infinitely long open string with two "arms" in the
asymptotically flat regions and a turning point {\it outside} the
ergosphere
\cite{fro}. The perturbation equations (3.20) with the potentials
(3.39)-(3.40)
then determine the reflection and transmission of waves (`phonons')
travelling along the string between the
asymptotically flat regions \cite{all,all9}. However, in
the special case $|b|=|a|,$ the string passes the static
limit and spirals inside the ergosphere towards the horizon. This is
the case
we are interested in, c.f. the discussion after eq.(3.5).
Notice that the potential (3.40), in general being divergent at the
static limit $(\Delta=a^2),$ is finite for $|b|=|a|,$ i.e. the
divergences precisely cancel out in the particular case where the
string actually crosses the static limit.

Let us now consider the case $|b|=|a|$ in more detail.
We use the notations $\Sigma_{\pm}$ for a pair of
string configurations connected by the reflection
transformation, $\phi\rightarrow
-\phi,$ discussed after eq.(3.36).
In order to make this prescription unique we choose
$dr/d\phi < 0$ for $\Sigma_+$ and $dr/d\phi > 0$ for $\Sigma_-$.
Then we find from eq.(3.11):
\begin{equation}
\hat{\sigma}=r,\;\;\;\;\;\;\;\;\;\;\;\hat{\tau}=t\pm
a^2\int^r\;\frac{2Mr
-Q^2}
{\Delta(\Delta-a^2)}\;dr,
\end{equation}
where the signs $\pm$ correspond to the string configuration
$\Sigma_{\pm}$. For both  configurations $\Sigma_{\pm}:$
\begin{equation}
\hspace*{-6mm}F=1-\frac{2M}{r}+\frac{Q^2}{r^2},
\end{equation}
thus the world-sheet line element (3.5) takes the form of a
2-dimensional
Reissner-Nordstr\"{o}m black hole. The parameters of mass $M$
and charge $Q$ of this 2-dimensional black hole are the same
as the corresponding  parameters of the original 4-dimensional
Kerr-Newman
metric.  The world-sheet spatial   coordinate
$\hat{\sigma}$ equals the Boyer-Lindquist radial coordinate $r,$
while
the
world-sheet time $\hat{\tau}$ approaches the Boyer-Lindquist time $t$
in the
asymptotically flat region, $r\rightarrow\infty.$  The coordinates
$(\hat{\tau}, r)$ cover only the exterior region of the string hole.
One can easily obtain the string metric valid in a wider region.
For this purpose it is convenient to introduce  the
Eddington-Finkelstein coordinates
$(u_{\pm},\varphi_{\pm}):$
\begin{equation}\label{EF}
dt=du_{\pm}\mp \Delta^{-1} (r^2+a^2)dr,\hspace{1cm}
d\phi=d\varphi_{\pm}\mp \Delta^{-1}a dr ,
\end{equation}
and  to rewrite the Boyer-Lindquist metric (\ref{BL}) as \cite{MTW}:
\begin{eqnarray}\label{EF1}
ds^2\hspace*{-2mm}&=&\hspace*{-2mm}
-\frac{\Delta}{\rho^2}[du_{\pm} -a\sin^2\theta d\varphi
_{\pm}]^2
+\frac{\sin^2\theta}{\rho^2}
[(r^2+a^2)d\varphi_{\pm} -a du_{\pm}]^2\nonumber \\
\hspace*{-2mm}&+&\hspace*{-2mm}
\rho^2 d\theta^2 \pm 2dr [du_{\pm} -a\sin^2\theta d\varphi_{\pm} ].
\end{eqnarray}
In these coordinates, the strings $\Sigma_{\pm}$ are described by
equations $\theta=\pi/2$, $\varphi_{\pm}=\mbox{const.},$ so that the
induced metric on $\Sigma_{\pm}$ is:
\begin{equation}\label{2EF}
ds^2=-Fdu_{\pm}^2\pm 2dr du_{\pm}.
\end{equation}
This metric  for $\Sigma_+$ describes a black hole, while  for
$\Sigma_-$
it describes a white hole.
In both cases the perturbation  equations (3.20) on $\Sigma_{\pm}$
reduce to:
\begin{equation}
\Box\Phi^R=\frac{\Box r}{r}\Phi^R;\;\;\;\;\;\;\;\;R={\tiny{
\perp,\;\parallel}}
\end{equation}
where $\Box$ is the d'Alambertian on the world-sheet,
$\Box=G^{AB}\nabla_A
\nabla_B$. We also have $\Box r= F_{,r}$.

Several interesting remarks are now in order. First notice that the
perturbation equations for the two transverse polarizations $\Phi^R$
are decoupled
and {\it identical}. Secondly, eq.(3.46) is precisely the $s$-wave
scalar field equation in the 4-dimensional Reissner-Nordstr\"{o}m
black
hole background:
\begin{equation}
\Box^{(4)}\phi=0,\;\;\;\;\;\;g_{\mu\nu}^{(4)}=
\mbox{diag}(-F,\;1/F,\;r^2,\;r^2\sin^2\theta),
\end{equation}
with $F$ given by eq.(3.42).
The decomposition $\phi=\sum r^{-1}\Phi_l (r,t)\;Y_{lm}(\theta,\phi)$
yields:
\begin{equation}
-\frac{1}{F}\partial_t^2\Phi_l+\partial_r(F\partial_r\Phi_l)=
\frac{F_{,r}}{r}\Phi_l+
\frac{l(l+1)}{r^2}\Phi_l,
\end{equation}
which is identical to equation (3.46) for $l=0,$ as is most easily seen
by writing eq.(3.46) in the $(\hat{\tau},r)$ world-sheet coordinate
system.
(Equation (3.48) has
been studied
in detail in the literature, see for instance \cite{chan} and
references
given therein.)
The string exitations for a stationary string passing through the
ergosphere in the equatorial plane
of a Kerr-Newman black hole are therefore described mathematically
by the $l=0$ scalar waves in the background of a 4-dimensional
Reissner-Nordstr\"{o}m black hole.

The above results allow  generalization to the case when a
stationary string
is located not in the equatorial plane but on the cone
$\theta=\theta_0 \ne \pi/2$.
For this case the parameters which enter the equations (3.36)
are related as $q^2=2|ab|$. The corresponding $\theta_0$ is
determined
as the minimum of the potential $U(\theta)$ and is, $\sin ^2
\theta_0=|b/a|$.
This relation implies that $|b|<|a|$.
The remarkable fact is that such a string allows a simple geometrical
description. The Kerr-Newman metric possesses two
principal null geodesic congruences, one of them is formed by
incoming and the other  by outgoing principal null rays \cite{MTW}.
Take one of these null geodesics $\gamma_{\pm}$ ($-$ stands for
an outgoing and $+$ stands for an incoming ray). Consider
two-dimensional
surfaces $\Sigma_{\pm}$ formed by Killing trajectories passing
through $\gamma_{\pm}$. It is possible to prove that $\Sigma_{\pm}$
is a minimal surface and it describes an "equilibrium" string
configuration
with the parameter $q^2=2|ab|.$
Two string configurations $\Sigma_{\pm}$
differ by signs of $r'$ in (3.36). The metric induced on both
surfaces $\Sigma_{\pm}$ possesses the Killing horizon. In case of
$\Sigma_+$ it describes a black hole, while in case of $\Sigma_-$
it describes a white hole (for a white hole a future directed
timelike
curve crossing  the Killing horizon enters the black hole exterior).
The above considered case of a stationary string located
on the equatorial plane and crossing
the infinite-red-shift surface is a special example of stationary
cone strings. The perturbation equations for cone strings have been
considered in \cite{hendy}.

To summarize, we have shown that the metric induced on a stationary
string crossing the infinite-red-shift surface  describes a
two-dimensional
geometry of a black or white hole. It opens remarkable possibilities
to  apply results of
mathematical study of two-dimensional black holes to
physical objects  (cosmic strings in the vicinity of a black hole),
which at least in principle allow experimental observations.
In particular, in the presence of the horizon on $\Sigma_+$
(for the two-dimensional string black hole) the conditions that the
quantum state is regular near the horizon implies that
 the string perturbations $\Phi^R$ are to be
thermally excited. It means that there will be a thermal flux of the
string
excitations ('phonons') propagating to infinity which forms the
corresponding
Hawking radiation. For a radial string crossing the event horizon
of the Schwarzschild black hole this effect was considered in
\cite{law}. We would like to stress that the analogous radiation
will
also be present  when a stationary string crosses the
infinite-red-shift surface (which for a rotating black hole is
located outside
the horizon).
String perturbations ('phonons') generated in
the region lying inside the 2-horizon and propagating along the
string $\Sigma_+$
cannot escape to infinity.
But it is well known that the causal signals emitted in the
ergosphere
and propagating in the 4-dimensional space-time can reach a
distant observer. One can use such signals ('photons') in order to
get information from the interior of the two-dimensional string
black  hole.
This situation is similar to one which happens in  a
'dumb' hole considered by Unruh \cite{Unruh}.  In order to define
a 'dumb' hole one uses the causal structure connected with
the propagation of phonons. The photons propagating with
supersonic velocity can escape a 'dumb' hole interior.
The nice feature which differs our model is that it is constructed
in the framework of completely relativistic theory.

Possibility of getting information from a black hole interior was
also
discussed in \cite{FrNo:93}, where  a gedanken experiment was
proposed in which  a traversable wormhole is used.
The mechanism discusses in the present paper which makes it
possible to extract
information from the interior of string black holes is different.
It is connected with the presence of extra-dimensions and does
not require
non-trivial topology.
The possibility of the information extraction from the interior of
a string black hole might give new insight to the problem of
information loss in black holes.
In particular the arguments of  Ref.\cite{Gott:95}  being applied to
string black holes indicate that black hole complementarity
may be inconsistent, at least for these black hole models.
\section*{Acknowledgements}
\setcounter{equation}{0}
 I thank N. S\'{a}nchez and H.J. de Vega
as well as V. Frolov
and S. Hendy for collaboration on different parts of the
material presented in this talk.
I also acknowledge the financial support from NSERC and
from NATO Collaboration Grant CRG 941287.

\newpage
\begin{centerline}
{\bf Figure Captions}
\end{centerline}
\vskip 24pt
\hspace*{-6mm}Figure 1. The potential $V(r),$ eq.(2.7), for
a circular string in the
equatorial plane of the four $3+1$ dimensional spacetimes: (a) Minkowski,
(b) Schwarzschild black hole, (c) anti de Sitter and (d) Schwarzschild
anti de Sitter. The details are explained in the text.
\vskip 12pt
\hspace*{-6mm}Figure 2. The potential $V(r),$ eq.(2.7), for
a circular string in the
equatorial plane of the two $3+1$ dimensional spacetimes: (a) de Sitter
and (b) Kerr black hole. The details are explained in the text.
\vskip 12pt
\hspace*{-6mm}Figure 3. The potential $U(r),$ eq.(2.13), for
a stationary string in the
equatorial plane of the three $3+1$ dimensional spacetimes: (a) Minkowski,
(b) anti de Sitter and (c) Schwarzschild black hole.
The details are explained in the text.
\vskip 12pt
\hspace*{-6mm}Figure 4. The potential $U(r),$ eq.(2.13), for
a stationary string in the
equatorial plane of de Sitter spacetime. The details are explained in the text.
\vskip 12pt
\hspace*{-6mm}Figure 5. The so-called
$(N,M)=(3,2)$ stationary string solution inside the
horizon of de Sitter
spacetime. Besides the circular string, this is the simplest stationary
closed string configuration in de Sitter spacetime.
\end{document}